\documentclass[12pt,preprint]{aastex}

\shorttitle{Interaction between magnetic fields}
\shortauthors{Zhang et al.}

\begin{document}

\title{Interaction between a fast rotating sunspot and ephemeral regions
associated with the major solar event on 2006 December 13}

\author{Jun Zhang\altaffilmark{}, Leping Li\altaffilmark{},
and Qiao Song \altaffilmark{}}

\altaffiltext{}{National Astronomical Observatories, Chinese
Academy of Sciences, Beijing 100012, China;
zjun@ourstar.bao.ac.cn, lepingli@ourstar.bao.ac.cn,
qiaosong@ourstar.bao.ac.cn}

\begin{abstract}

The major solar event on 2006 December 13 is characterized by the
approximately simultaneous occurrence of a heap of hot ejecta, a
great two-ribbon flare and an extended Earth-directed coronal mass
ejection. We examine the magnetic field and sunspot evolution in
active region NOAA AR 10930, the source region of the event, while
it transited the solar disk centre from Dec. 10 to Dec. 13. We find
that the obvious changes in the active region associated with the
event are the development of magnetic shear, the appearance of
ephemeral regions and fast rotation of a smaller sunspot. Around the
area of the magnetic neutral line of the active region, interaction
between the fast rotating sunspot and the ephemeral regions triggers
continual brightening and finally the major flare. It is indicative
that only after the sunspot rotates up to 200$^{\circ}$ does the
major event take place. The sunspot rotates at least 240$^{\circ}$
about its centre, the largest sunspot rotation angle which has been
reported.

\end{abstract}


\section{INTRODUCTION}

Solar flares are powered by the energy stored in the stressed
magnetic field, strong flares tend to occur in the vicinity of
magnetic neutral lines where the field gradients are strong and the
horizontal components are highly sheared \citep[e.g.,][]{har76,
wan94, wan02, den06}. The movement of magnetic footpoints by
photospheric flows can lead to the destabilization of the magnetic
field and hence to flares by increasing the length of the field
lines in the corona \citep[e.g.,][]{som02}. On the other hand,
\citet{nin02} have pointed out that shearing motions have little
effect in the process of build-up of magnetic free energy that leads
to the initiation of coronal mass ejections (CMEs).

Observations indicate that reconnection-favored emerging flux has a
strong correlation with flare onset and CMEs \citep{sch97, ish98,
kus02, kur02, sak04}. Based on the flux rope model, an emerging flux
trigger mechanism is proposed for the onset of CMEs, using
two-dimensional magnetohydrodynamic (MHD) numerical simulations
\citep{che00}.

Besides magnetic field observations in active regions, white-light
observations, furthermore, have shown sunspot evolution, e.g.
sunspot rotation \citep{eve10, mal64, gop65}. \citet{ste69} has
suggested that sunspot rotation may be involved with energy build-up
and later release by a flare. With the high spatial and temporal
resolution of recent satellite-borne telescopes, the observations of
rotating sunspots and other magnetic structures have become more
frequent and easier to identify \citep{nig00, nig02, bro01}. The
sigmoid structures \citep{can99}, which are thought to be more
likely to erupt and cause flares and CMEs, may be associated with
the rotation of the magnetic footpoints in the photosphere.
\citet{bro03} have shown that some sunspots rotate up to
200$^{\circ}$ about their umbral centre, and the corresponding loops
in the coronal fan twist and erupt as flares.

The major event on 2006 December 13 exhibits almost simultaneous
plasma ejecta, flare activity and a CME. It provides a good
opportunity to study the surface magnetic activity and sunspot
kinematics that results in the rather global magnetic instability.
In this Letter, we examine the magnetic evolution and sunspot
rotation prior to and during the course of the major event with the
emphasis on the interaction of a rotating sunspot and ephemeral
regions that characterizes this CME-producing active region.

\section{OBSERVATIONS}

Active region NOAA AR 10930 displayed the evolution of the complex
magnetic structure, surrounded by many small dark pores and bright
magnetic faculae, as observed by the {\it SOHO}/ MDI \citep{sch95}
and the Transition Region and Coronal Explorer \citep[{\it TRACE},]
[]{han99} over a 12-day interval from 2006 December 6 to December
18. The data set analyzed consists of four-day (from Dec. 10 to 13)
full-disk magnetograms obtained from MDI, synchronous
high-resolution white-light and UV (1600 {\AA}) observations from
the {\it TRACE} satellite, with spatial resolution of 1.0$''$,
temporal resolution of 20 - 60 s, and field-of-view of
384$''$${\times}$384$''$. Fig. 1 shows general appearance of the
active region NOAA AR 10930 in the decaying phase of the major
flare.

\section{CONCLUSIONS AND DISCUSSIONS}

By checking the MDI and {\it TRACE} data, we notice that the first
obvious magnetic evolution is the developing of magnetic shear. Fig.
2 presents the time sequence of the MDI longitudinal magnetograms
(left), the corresponding {\it TRACE} continuum images (middle) and
the corresponding {\it TRACE} 1600 {\AA} images (right). Two
contours in the continuum images at Dec. 10, 12:50 UT and 22:30 UT
outline a dark thread, which connects the pair of opposite polarity
sunspots of the active region. These contours are overplotted onto
the corresponding longitudinal magnetograms and {\it TRACE} 1600
{\AA} images. Before Dec. 10, 12:51 UT, the thread connects directly
with the two sunspots. Ten hours later, although the thread still
joins the two sunspots, its shape changes from beeline to reverse
S-shape curve. Gradually the thread disappears and some new threads
appear along the neutral line of the two sunspots (see the dash
curves on the continuum image in the low panel). From the MDI
magnetogram on Dec. 11 at 12:51 UT, we find that these new threads
seem to be arch filament systems connecting the opposite polarities
of ephemeral regions (ERs) which emerge along the sunspot neutral
line. Fig. 3 clearly shows an ER (denoted by two pairs of arrows) on
high resolution MDI magnetograms, after the magnetic shear of the
active region is well developed. Firstly the two magnetic elements
belonging to the ER are ellipses in shape, and separate each other
with a speed of 0.5 km s$^{-1}$. After four-hour evolution, the
shape of the elements looks like two narrow ribbons along the
neutral line of the active region, as the ER is squeezed tightly by
the region. In the process of the ER appearing, multiple neutral
lines (marked by dark lines in the 02:40 UT magnetogram) develop,
and continuous UV brightening appears in the area of these neutral
lines.

We have noticed that ERs appear not only prior to the onset of the
major solar event, but also during the course of the event. Fig. 4
shows an ER appearing during the flare/CME event. The arrows point
to the ER. The positive element of the ER plunges into the neutral
line area of the active region. Before it merges to the positive
polarity magnetic field of the active region, it cancels with
surrounding opposite polarity field, meanwhile UV brightening
appears. The bright material ejects along the axis of the ER to the
right (shown by the arrow in the {\it TRACE} 1600 {\AA} image at
01:40 UT). Twenty minutes later, the major flare takes place. The
negative element moves away from the active region with a speed of
0.5 km s$^{-1}$. At 06:05 UT, the distance between the two elements
is about 22,000 km, and an UV bright ribbon connects with the two
elements of the ER.

Observations from {\it TRACE} in the photospheric white-light
channel have shown that, from the end of Dec. 10 on, sunspot
rotation in the active region accompanies the magnetic flux
emergence in the form of ERs in the neutral line area. Fig. 5 shows
an example of the rotation of a dark penumbral feature `f3' (marked
by the arrows). The feature firstly appeared near Dec. 12, 00:10 UT,
it rotated around the centre (shown by smaller circles) of the
smaller spot. From Dec. 12, 00:10 UT to 20:50 UT, the dark feature
rotated about 190$^{\circ}$, and the mean angular speed reaches
9$^{\circ}$ hr$^{-1}$. In the following 14 hr, the feature became
larger and larger, due to the converging of other unresolved
features. The angular speed decreased to 2$^{\circ}$ hr$^{-1}$ on
average. Finally the dark feature broke up and gradually decayed. By
checking the continuum data, we can clearly identify that in the
penumbra of the smaller sunspot, there are several dark features
which rotated around the spot. Fig. 6 shows the rotation of the
three penumbral features mentioned in Fig. 5. It shows that a
penumbral feature started to rotate while the shear developed, and
the rotational speed is about 20$^{\circ}$ hr$^{-1}$. Between 02:00
UT and 12:00 UT on Dec. 12, the rotational speed of all the three
features is low with a mean value of 4$^{\circ}$ hr$^{-1}$. Several
hours prior to the flaring activity, all the three features
underwent fast rotational process (see the small circles in the
bottom panel). Before the flare took place, the rotational angle of
each feature exceeded 200$^{\circ}$. During the course of the
flaring activity, these features displayed a relatively low
rotational speed. After the onset of the flaring activity, these
features continuously rotated around the sunspot with a mean speed
of 2$^{\circ}$ hr$^{-1}$. Near the end of Dec. 13, the amount of the
rotational angle for each feature was around 240$^{\circ}$. This is
the largest angle which has been reported.

The major solar event manifests itself as a heap of bright ejecta, a
great flare, and an extended Earth-directed CME. For such a major
event, the obvious activities in its source region are the
development of magnetic shear, the appearance of ephemeral regions
and fast rotation of a sunspot. More importantly, the major event
takes place only after the rotational angle of the sunspot reached
200$^{\circ}$. It is prefigurative that the interaction between
ephemeral regions and fast rotation of a sunspot plays a decisive
role in producing the global instability responsible for this major
solar event.

\citet{reg06} have identified that in active region NOAA AR 8210, a
fast motion of an emerging polarity is associated with small-scale
reconnections, and a sunspot rotation enables the occurrence of
flare by a reconnection process close to a magnetic surface.
\citet{wan93} presented a two-step reconnection scenario for flare
process. The first step of reconnection is seen as flux cancellation
observed in the photosphere \citep{zha01}. This reconnection
transports the magnetic energy and complexity into coronal magnetic
structure. The second step of reconnection is directly responsible
for transient solar activities, and takes place only when some
critical status is achieved in the corona.

By performing MHD simulations in which the twisting motions are
included, several authors \citep[e.g.,][]{mik90, gal97} have
reported that photospheric flows will drive the loop unstable to the
ideal kink instability after the average twist exceeds a critical
value of 450$^{\circ}$. Later, \citet{ger03} have simulated the
rotation of a pore around a sunspot. While the pore rotates
180$^{\circ}$ around the sunspot, the current increases rapidly as
the centre of the pore makes contact with the large sunspot. This
current build-up could be an explanation of an observed flaring.
Observational result in this Letter (e.g., see Fig. 6) is consistent
with the simulation of \citet{ger03}.

We are keenly aware of the limitations imposed by the
comparatively low spatial resolution of MDI magnetograms. Future
observations at higher spatial resolution are likely to uncover
the nature of the source regions of major solar events. A detailed
study of this flare/CME event is planned with Hinode data.

\acknowledgments

The authors are indebted to the {\it TRACE} and {\it SOHO}/MDI
teams for providing the data. {\it SOHO} is a project of
international cooperation between ESA and NASA. This work is
supported by the National Natural Science Foundations of China
(G10573025 and 40674081), the CAS Project KJCX2-YW-T04, and the
National Basic Research Program of China under grant
G2006CB806303.

\clearpage

\begin{figure}
\epsscale{0.90}
\caption{General appearance of the
active region NOAA AR 10930 on different wavelength in the decaying
phase of the major flare. {\it Upper left:} an MDI longitudinal
magnetogram. The window outlines a subarea where ephemeral regions
appear (see Fig. 4); {\it Lower left:} A continuum intensity image
from {\it TRACE}. `N1' is the main sunspot of the active region,
`P1' and `P2' are the following sunspots. The window denotes the
field-of-view of Fig. 5 where the sunspot `P1' fast rotates; {\it
Upper right:} a {\it TRACE} 1600 {\AA} image showing the two ribbons
(`R1' and `R2') of the flare; {\it Lower right:}  a {\it TRACE} 195
{\AA} image showing the post-flare loops. The field-of-view is about
100$''$${\times}$100$''$. \label{fig1}}
\end{figure}

\clearpage

\begin{figure}
\epsscale{0.90}
\caption{The time sequence of the
MDI longitudinal magnetograms (the left column), the corresponding
{\it TRACE} continuum images (the middle column) and the
corresponding {\it TRACE} 1600 {\AA} images (the right column),
showing the developing of magnetic shear. The field-of-view is about
50$''$${\times}$50$''$. The dotted contours and curves are described
in the text. \label{fig2}}
\end{figure}

\clearpage

\begin{figure}
\epsscale{0.80}
\caption{Time sequence of the high resolution MDI longitudinal
magnetograms, showing the emergence flux appeared near the magnetic
neutral line of the active region, after the AR 10930 magnetic shear
is well developed. The field-of-view is about
60$''$${\times}$60$''$. The arrows and the dark lines are described
in the text. \label{fig3}}
\end{figure}

\begin{figure}
\epsscale{0.60}
\caption{Time sequence of the MDI
longitudinal magnetograms (the left column), the corresponding {\it
TRACE} continuum images (the middle column) and the corresponding
{\it TRACE} 1600 {\AA} images (the right column), showing the
appearance of bipolar magnetic features during the course of the
major flare/CME event. The field-of-view is about
50$''$${\times}$50$''$. The arrows are described in the text.
\label{fig4}}
\end{figure}

\begin{figure}
\epsscale{0.90}
\caption{Time sequence of {\it
TRACE} continuum images showing the rotation of a dark penumbral
feature (`f3') around the centre (marked by small circles) of `P1'
mentioned in Fig. 1. `f1' and `f2' are other two rotating dark
features which appear prior to the emergence of `f3'. The three
arrows in the continuum image at Dec. 12, 09:12 UT point to the
three features (`f1', `f2' and `f3'), otherwise the arrows points to
`f3'. The convergence of `f1', `f2' and other unresolved features
forms `P2' (see Fig. 1). The two solid lines in the first and last
images connect `f3' with the centre-of-gravity of `P1', and the
dotted line in the last image is a duplicate of the solid line in
the first image. The field-of-view is about 40$''$${\times}$40$''$.
\label{fig5}}
\end{figure}

\begin{figure}
\epsscale{1.0}
\vspace{-0.5cm}
\caption{Plots
showing rotational angle (top) of the three penumbral features
(`f1', `f2' and `f3' shown in Fig. 5) and rotational speed (bottom)
of the corresponding features versus time. Dark gray area represents
the shear developing period, and light gray area the flaring
activity period. The minus rotation of these features is caused by
their moving away from the sunspot. \label{fig6}}
\end{figure}
\clearpage

\end{document}